\documentclass[aps, prd, twocolumn, superscriptaddress, nofootinbib]{revtex4}
\usepackage{eurosym}
\usepackage{graphicx}
\usepackage{dcolumn}
\usepackage{amssymb}
\usepackage{amsmath,amssymb,amsfonts}

\newcommand{\be}{\begin{equation}}
\newcommand{\ee}{\end{equation}}
\newcommand{\bea}{\begin{eqnarray}}
\newcommand{\eea}{\end{eqnarray}}

\begin{document}

\title{Redshifting of cosmological black bodies in BSBM varying-alpha
theories}
\author{John D. Barrow}
\affiliation{DAMTP, Centre for Mathematical Sciences, University of Cambridge,
Wilberforce Rd., Cambridge, CB3 0WA, U.K.}
\author{Jo\~{a}o Magueijo}
\affiliation{Theoretical Physics, Blackett Laboratory, Imperial College, London, SW7 2BZ,
U.K.}
\affiliation{Dipartimento di Fisica, Universit\`{a} La Sapienza and Sez. Roma1 INFN, P.
le A. Moro 2, 00185 Rome, Italy}
\date{\today }

\begin{abstract}
We analyse the behaviour of black-body radiation in theories of
electromagnetism which allow the electron charge and the fine structure
constant to vary in space and time. We show that such theories can be
expressed as relativistic generalizations of a conventional dielectric. By
making the appropriate definition of the vector potential and associated
gauge transformations, we can identify the equivalent of the electric and
displacement fields, $\mathbf{E}$ and $\mathbf{D}$, as well as the magnetic $%
\mathbf{B}$ and $\mathbf{H}$ fields. We study the impact of such dielectrics
on the propagation of light in the so-called \textquotedblleft
BSBM\textquotedblright\ theory. We examine the form of simple cosmological
solutions and conclude that no changes are created to the standard
cosmological evolution of the temperature and energy-density of black-body
radiation. Nonetheless the matter evolution changes and the behaviour of the
entropy per baryon is modified, and the ratios of different dark matter
components may be changed too.
\end{abstract}

\maketitle


\section{Introduction}

There has been considerable observational and theoretical interest in the
possibility that some dimensionless atomic constants of Nature, like the
fine structure constant, $\alpha ,$ or the proton-electron mass ratio, $\mu $%
, might be varying very slowly in space and time \cite{con}. A range of
high-precision astronomical instruments have opened up new ways to find very
small changes in the values of physical \textquotedblleft
constants\textquotedblright\ that would be undetectable in current
terrestrial experiments \cite{obs}. Often, observational data is used simply
to deduce the value of a constant at some redshift $z>0,$ and compare it
with the value measured here and now in the laboratory at $z=0$: no actual
theory of the constant's variation is used to connect the two values or to
include other possible consequences of varying constants on the cosmological
history. By contrast, theories which promote constants to become variables
in a self-consistent way do so by making them into spacetime variable
(scalar) fields. These fields must then gravitate and satisfy the
constraints imposed by energy and momentum conservation. These requirements
determine the generalisations of Einstein's general relativity which
incorporate varying constants.

In the well-studied case of varying $G,$ if we ignored this dictate and
simply compared values of $G$ at fixed $z,$ then the conclusions drawn would
be unreliable because the time-variation of $G$ couples to the evolution of
the cosmological scale factor at zero-order. A self-consistent Brans-Dicke
(BD) scalar-tensor theory of gravity \cite{bd} has a solution for a
time-variation of the form $G\propto t^{-n}$ and a cosmological scale factor
evolution $a(t)\propto t^{(2-n)/3}$ in a flat, dust-filled BD universe \cite%
{lid} (and its Newtonian analogue has the same solution \cite{jbnewt}). In
the case of a varying $\alpha ,$ a corresponding self-consistent scalar
theory for its space-time variation is provided by the theory of Bekenstein 
\cite{bek}, Sandvik, Barrow and Magueijo (BSBM) \cite{sbm1, sbm2, Blip, bmot}%
. Exact, approximate and asymptotic solutions of this theory and its
extensions can be found and used to fit observational data from quasar
spectra and elsewhere. Analogous theories have also been created to study
self-consistently the variation of $\mu $ \cite{bm1}. Unlike the situation
with varying-$G,$ variations of atomic \textquotedblleft
constants\textquotedblright , like $\alpha $ and $\mu ,$ do not have
significant effects on the evolution of the cosmological scale factor; for
example during a cold-dark-matter era $\alpha (t)\propto \ln (t/t_{0})$ and $%
a(t)\propto t^{2/3}[\ln (t/t_{0})]^{\lambda }$ where $\lambda \simeq 3\times
10^{-5}$ and $t_{0}$ are constants \cite{bmot}.

Despite this general pattern, it has been claimed that variations of $\alpha 
$ and $\mu $ are able to create discernible (logarithmic) differences in the
temperature-redshift evolution for massless particles in BSBM theories \cite%
{mar1, bg, mar2}. This is one aspect of the original formulation of a
varying $\alpha $ theory by BSBM that will be the focus of this paper. In
the original formulation it appears that the evolution of a black body
distribution of equilibrium photons can pick up logarithmic corrections to
the standard cosmological evolution of the standard temperature-redshift
relation, $T_{\gamma }\propto a^{-1}\propto (1+z)$, changing it to $%
T_{\gamma }\propto \alpha ^{1/4}a^{-1}\propto \alpha ^{1/4}(1+z)$. If true,
this would have potentially observable consequences -- including slow
evolution of the photon entropy per baryon and of the ratio of the neutrino
to photon temperatures, and deviation from linearity in the relation between
radiation temperature and redshift $z$. All have observational consequences.
In this paper we analyse this feature of BSBM theories in detail and show
that in BSBM, with the appropriate choice of generalisation from Maxwell's
theory, the evolution of black-body radiation follows the standard
cosmological trajectory followed by theories with constant $\alpha $. The
same conclusions will hold \textit{mutatis mutandis} for theories of this
type with varying $\mu $ \cite{bm1}.

In order to investigate how a black body reacts to the BSBM field, $\psi $,
and whether it might self-consistently drive variations in the electron
charge $e\equiv e_{0}\exp (2\psi )$ (and hence in $\alpha \equiv
e^{2}/\hslash c$), we develop an analogy with a dielectric medium \cite{LL}
in Section~\ref{diel}. We find that the BSBM field $\psi $ behaves like a
relativistic generalization of a dielectric or insulator. It is linear ($%
\mathbf{D}$ and $\mathbf{H}$ are proportional to $\mathbf{E}$ and $\mathbf{B}
$) and the proportionality constants $\epsilon $ and $\mu ^{-1}$ are
isotropic and frequency independent. Unlike in standard media, $\epsilon $
and $\mu ^{-1}$ obey a relativistic Klein-Gordon equation that is sourced by
the EM lagrangian. We find that $\epsilon =1/\mu $, so the medium is
non-dispersive. In Section~\ref{freeflight} we examine the properties of
photons in ' free'\ flight through such a dielectric and find no frequency
shift or photon production. Hence we do not expect a non-interacting black
body to be affected by changes in $\alpha $ created by the space-time
variation of $\psi $. The same should hold true for a coupled thermalised
system. Therefore there will be no new observable effects on the redshift
history of the cosmic microwave background radiation.

\section{BSBM varying-alpha as a relativistic dielectric effect}

\label{diel} BSBM \cite{bek,sbm1,sbm2} was built upon the principles of
relativistic Lorentz invariance and the gauge principle. The $\psi $ field
it employs to carry variations in $\alpha $, which obeys a relativistic
Klein-Gordon equation, can therefore never be identical to a dielectric
medium in conventional electrodynamics, which is usually a non-relativistic
material (an insulator). However, the $\psi $ field may be regarded as a
relativistic generalization of a conventional dielectric. Although this
creates some important differences, nevertheless with regard to the effects
on electromagnetic radiation much of the formalism is similar, as we now
show.

In setting up BSBM there is an ``ambiguity'' in the definition of electric
and magnetic fields similar to that found for insulators, where one can use $%
\mathbf{E}$ or $\mathbf{D}$ for the electric field, and $\mathbf{B}$ or $%
\mathbf{H}$ for the magnetic field. In reality both concepts play a role,
with $\mathbf{E}$ and $\mathbf{B}$ convenient for writing the homogeneous
Maxwell equations, and $\mathbf{D}$ and $\mathbf{H}$ better suited for
writing the inhomogeneous equations, even when there are no sources.

A first decision fork appears in the literature in the definition of the
vector potential (and the expression of gauge transformations). One can use
either $A_{\mu }$ (as in \cite{bek}), or $a_{\mu }$ (as in \cite{sbm1}),
with the two related by: 
\begin{equation}
a_{\mu }=e^{\psi }A_{\mu }\,,
\end{equation}%
where 
\begin{equation}
\psi =\ln \tilde{\epsilon}=\ln \frac{e}{e_{0}}=\frac{1}{2}\ln \alpha .
\end{equation}%
The last expression links $\psi $ to the fine structure ``constant'', $%
\alpha $. Here $\tilde{\epsilon}$ corresponds to the \textquotedblleft $%
\epsilon $\textquotedblright\ used in \cite{bek}, which we stress is \textit{%
not} the relative permittivity of the \textquotedblleft
medium\textquotedblright ($\epsilon $,\ in our notation here), as we shall
see. Gauge transformations can be performed as 
\begin{equation}
a_{\mu }\rightarrow a_{\mu }+\partial _{\mu }\Lambda
\end{equation}
or as 
\begin{equation}
A_{\mu }\rightarrow A_{\mu }+\frac{\partial _{\mu }\Lambda }{\tilde{\epsilon}%
}.
\end{equation}
This fork propagates into the definition of gauge-invariant field tensors,
with \cite{bek} led to the natural definition: 
\begin{equation}  \label{Fdef}
F_{\mu \nu }=e^{-\psi }\left[ \partial _{\mu }(e^{\psi }A_{\nu })-\partial
_{\nu }(e^{\psi }A_{\mu })\right] ,
\end{equation}%
and \cite{sbm1} to 
\begin{equation}
f_{\mu \nu }=\partial _{\mu }a_{\nu }-\partial _{\nu }a_{\mu }.  \label{fdef}
\end{equation}%
The two are related by 
\begin{equation}
F_{\mu \nu }=e^{-\psi }f_{\mu \nu }.
\end{equation}%
The electromagnetic action, from which the non-homogeneous Maxwell's
equations are derived, can be written in the two forms: 
\begin{equation}  \label{action}
S_{EM}=-\frac{1}{4}\int d^{4}x\,F^{2}=-\frac{1}{4}\int d^{4}x\,e^{-2\psi
}f^{2}.
\end{equation}

In order to study which quantities play the role of $\mathbf{E}$ and $%
\mathbf{B}$ we examine the non-homogeneous Maxwell equations. These are best
written in terms of $f_{\mu \nu }$, in the form of the integrability
condition: 
\begin{equation}
\epsilon ^{\alpha \beta \mu \nu }\partial _{\beta }f_{\mu \nu }=0.
\end{equation}%
This is obviously a necessary condition for (\ref{fdef}), but note that the
same argument cannot be made directly for $F_{\mu \nu }$ (derivatives of $%
\psi$ would appear in the corresponding condition in terms of $F_{\mu\nu}$;
cf. (\ref{Fdef}) and (\ref{fdef})). Thus, in order to parallel the usual
theory of electrodynamics in media we should associate $\mathbf{E}$ and $%
\mathbf{B}$ (appearing in the inhomogeneous Maxwell equations) with $f_{\mu
\nu }$, with entries in the usual places. With this identification we obtain
the standard inhomogeneous Maxwell equations: 
\begin{eqnarray}
\nabla \cdot {\mathbf{B}} &=&0, \\
\nabla \wedge {\mathbf{E}}+\frac{\partial {\mathbf{B}}}{\partial t} &=&0\;.
\end{eqnarray}%
This was already noted in \cite{bek} (however, the wrong identification was
made in ref.\cite{kraisel}, cf. their Eq.(25)).

In order to find the equivalent of $\mathbf{D}$ and $\mathbf{H,}$ we
consider instead the inhomogeneous Maxwell equations. In the absence of
currents these can be written in the two forms: 
\begin{equation}
\partial _{\mu }(e^{-2\psi }f^{\mu \nu })=\partial _{\mu }(e^{-\psi }F^{\mu
\nu })=0,
\end{equation}%
and we see that neither of them leads to the equivalent standard expression
for dielectric media (in both cases extra terms in the derivatives of $\psi $
appear). Therefore, we should define the alternative tensor: 
\begin{equation}
\mathcal{F}_{\mu \nu }=e^{-\psi }F_{\mu \nu }=e^{-2\psi }f_{\mu \nu },
\end{equation}%
in terms of which we have 
\begin{equation}
\partial _{\mu }\mathcal{F}^{\mu \nu }=0\,.
\end{equation}%
We should then define $\mathbf{D}$ and $\mathbf{H}$ from the appropriate
entries in $\mathcal{F}_{\mu \nu }$, so as to get: 
\begin{eqnarray}
\nabla \cdot {\mathbf{D}} &=&0, \\
\nabla \wedge {\mathbf{H}}-\frac{\partial {\mathbf{D}}}{\partial t} &=&0\;.
\end{eqnarray}%
With these identifications BSBM becomes equivalent to electromagnetism in
dielectric media with only small adaptations. We have: 
\begin{eqnarray}
\mathbf{D} &=&\epsilon \mathbf{E}=e^{-2\psi }\mathbf{E,} \\
\mathbf{H} &=&\mu ^{-1}\mathbf{B}=e^{-2\psi }\mathbf{B,}
\end{eqnarray}%
and so 
\begin{equation}
\epsilon =\frac{1}{\mu }=e^{-2\psi }.  \label{epsmu}
\end{equation}%
$\mathbf{D}$ and $\mathbf{H}$ are proportional to $\mathbf{E}$ and $\mathbf{B%
}$, and the proportionality constants $\epsilon $ and $\mu ^{-1}$ are
isotropic and frequency-independent. However, they do not depend locally on
the EM field (as is the case for standard media). Rather, they obey a
relativistic Klein-Gordon equation. Since $\epsilon =1/\mu $, the medium is
non-dispersive, as we shall explicitly prove in the next Section. Our
discussion has been for Minkowski spacetime, but replacing derivatives with
covariant derivatives gives the usual generalization to curved space-time 
\footnote{%
Several of these points were noted before in refs.\cite{dicke,bek}, although
our analogy of a relativistic dielectric medium was not used there. For
instance, in \cite{bek} (using the notation $\tilde{\epsilon}$\ for the
\textquotedblleft $\epsilon $\textquotedblright\ used therein), we should
have $\epsilon =1/\tilde{\epsilon}^{2}$, as recognised in the Section IIIC
of \cite{bek}.}.

\section{Electromagnetic waves: photons in free flight}

\label{freeflight} Maxwell's equations may be combined in the usual way to
produce electromagnetic wave equations. As in the case of conventional
dielectrics, these are most symmetrically written in terms of $\mathbf{E}$
and $\mathbf{H}$. For simplicity, assuming that $\epsilon =\epsilon (t)$ and 
$\mu =\mu (t)$ only, they are: 
\begin{eqnarray}
-\nabla ^{2}\mathbf{E}+\frac{\partial }{\partial t}\mu \frac{\partial }{%
\partial t}\epsilon \mathbf{E} &=&0, \\
-\nabla ^{2}\mathbf{H}+\frac{\partial }{\partial t}\epsilon \frac{\partial }{%
\partial t}\mu \mathbf{H} &=&0.
\end{eqnarray}%
Generalizations for space-dependent fields could be expressed, but will not
be required in this paper. So far we have not assumed that $n=1$.

\subsection{WKB solution}

A WKB expansion may now be sought, with ans\"{a}tze: 
\begin{eqnarray}
\mathbf{E} &=&\mathbf{e}_{E}E_{0}(t)\exp \left[ i\left( \mathbf{k}\cdot 
\mathbf{x}-\int \omega dt\right) \right] , \\
\mathbf{H} &=&\mathbf{e}_{H}H_{0}(t)\exp \left[ i\left( \mathbf{k}\cdot 
\mathbf{x}-\int \omega dt\right) \right] \;.
\end{eqnarray}%
To order $\omega ^{2},$ this translates into the dispersion relations: 
\begin{equation}
\omega ^{2}=n^{2}k^{2},
\end{equation}%
with refractive index 
\begin{equation}
n^{2}=\epsilon \mu \,.
\end{equation}%
To first order in $\omega $ we get the equations: 
\begin{eqnarray}
\frac{\dot{\mu}}{\mu }+2\frac{\dot{\epsilon}}{\epsilon }+2\frac{\dot{E}_{0}}{%
E_{0}} &=&0, \\
\frac{\dot{\epsilon}}{\epsilon }+2\frac{\dot{\mu}}{\mu }+2\frac{\dot{H}_{0}}{%
H_{0}} &=&0,
\end{eqnarray}%
so that 
\begin{eqnarray}
E_{0} &\propto &\frac{1}{\epsilon \sqrt{\mu }}, \\
H_{0} &\propto &\frac{1}{\mu \sqrt{\epsilon }},
\end{eqnarray}%
and also 
\begin{eqnarray}
D_{0} &\propto &\frac{1}{\sqrt{\mu }}, \\
B_{0} &\propto &\frac{1}{\sqrt{\epsilon }}\,.
\end{eqnarray}%
%
%
%
%
%
Specializing to BSBM, we can assume (\ref{epsmu}), and thus 
\begin{equation}
n=1,
\end{equation}%
so that BSBM does not induce modified dispersion relations or a variation in
the speed of light. This was noted before in \cite{bekschif}, and is hardly
surprising. It is built into the theory, since $\psi $ may be regarded as an
\textquotedblleft insulating aether\textquotedblright\ which does not break
Lorentz invariance. Therefore, it is expected that the central property of
light---the constancy of its speed---is left undisturbed, since it underpins
the Lorentz invariance.

The absence of dispersion in the free propagation of electromagnetic waves
implies that in BSBM there is no extra shift in the frequency of photons. In
addition, we can derive the scalings for the amplitudes: 
\begin{eqnarray}
E_{0} &\propto &e^{\psi }\propto B_{0}, \\
H_{0} &\propto &e^{-\psi }\propto D_{0}.
\end{eqnarray}%
These are consistent with $E_{0}=B_{0}$ and the orthogonality of $\mathbf{E}$
and $\mathbf{B}$.

\subsection{Energy density}

The stress-energy tensor in BSBM can be obtained taking variations of (\ref%
{action}) with respect to the metric. The expressions thus obtained match
the well-know results in the theory of electrodynamics in media. For
example, the energy density (i.e. $T^{00}$) is given by: 
\begin{equation}  \label{rho}
\widetilde{\rho} =\frac{1}{2}(ED+BH)=\frac{e^{-2\psi }}{2}(E^{2}+B^{2}).
\end{equation}%
For a wave packet, averaging over many wavelengths and periods, this
becomes: 
\begin{equation}
\widetilde{\rho} =\frac{1}{4}e^{-2\psi }(E_{0}^{2}+B_{0}^{2})\omega ^{2}=%
\frac{1}{2}e^{-2\psi }E_{0}^{2}\omega ^{2},
\end{equation}%
and so $\widetilde\rho$ does not depend on $\psi $, since $E_{0}\propto
B_{0}\propto e^{\psi }$, as we saw in our WKB solution.

Therefore, in BSBM the energy of a wave packet does not change because of a
varying $e$ (or $\alpha $). Combining this with the absence of dispersion we
can conclude that there is no photon production, since the density of
photons with frequency $\omega $ is $N=\rho /(\hbar \omega )$. The
Stefan-Boltzmann law for a decoupled black body is therefore also preserved,
with the temperature receiving no new effects due to a varying $\alpha $
(other than those due to the gravitational effect of $\psi $ in the Friedman
equation, which, as we stated, are negligible).

We note that if we want $\psi $ to be absent from the expressions of the
electromagnetic stress-energy tensor in terms of the fields, then, as
evident from (\ref{rho}), we should work with electric and magnetic fields
which are \textquotedblleft geometric averages\textquotedblright\ of $\ \ 
\mathbf{E}$ and $\mathbf{D,}$ and of $\mathbf{B}$ and $\mathbf{H}$. In fact,
these are the fields which make up Bekenstein's $F_{\mu \nu }$ (see \cite%
{bek}), and this is the tensor which enters without explicit coupling to $%
\psi $ in the Lagrangian and the stress-energy tensor (although of course $%
\psi $ is hidden inside the definition of $F_{\mu \nu }$ in terms of the
4-potential). We argue that this is the correct definition for the energy
for these theories, based on the results in this Section, as well as those
found in the next two.

\section{Cosmological equations}

\label{cosmo} The previous argument can be reinforced by looking at the
cosmological equations. With a standard Lagrangian for $\psi$: 
\begin{equation}
\mathcal{L}_\psi=-\frac{\omega_B}{2}\partial_\mu\psi\partial ^\mu\psi
\end{equation}
we are led to a driven Klein-Gordon equation for the evolution of $\psi$: 
\begin{equation}
\nabla ^2 \psi=\frac{2}{\omega_B} {\widetilde {\mathcal{L}}}_{EM}\, .
\end{equation}
where $\widetilde{\mathcal{L}}_{EM}=e^{-2\psi }(E^{2}-B^{2})/2$ (more
generally in what follows we denote as tilded variables those which have
absorbed factors which are a function of $\psi $ into their definition).
Under the assumption of homogeneity and isotropy, this Klein-Gordon becomes
the ODE: 
\begin{equation}
\ddot{\psi}+3\frac{\dot{a}}{a}\dot{\psi}=-\frac{2}{\omega_B }\widetilde{%
\mathcal{L}}_{EM}\,.  \label{KG}
\end{equation}%
and this equation can be interpreted as an energy balance equation, with the
driving terms representing energy exchange between $\psi$ and other forms of
matter. Indeed homogeneity and isotropy imply that $\psi $ must behave like
a perfect fluid, and computing the stress-energy tensor reveals: 
\begin{equation}
p_{\psi }=\rho _{\psi }=\omega_B \frac{\dot{\psi}^{2}}{2}.  \label{rhopsi}
\end{equation}
Equation (\ref{KG}) is then equivalent to: 
\begin{equation}  \label{dotrho}
\dot{\rho}_{\psi }+3\frac{\dot{a}}{a}(p_{\psi }+\rho _{\psi })=-2\dot{\psi}%
\widetilde{\mathcal{L}}_{EM}\;.
\end{equation}%
Each cosmic component $i$ contributes a term proportional to $\widetilde{ 
\mathcal{L}}_{EM_{i}}$ to the right hand side of (\ref{dotrho}). This should
be balanced by a counter-term with opposite sign in the right hand side of
the the conservation equation for $i$: 
\begin{equation}
\widetilde{\rho }_{i}+3\frac{\dot{a}}{a}(\widetilde{\rho }_{i}+\widetilde{p}%
_{i})=2\dot{\psi}\widetilde{\mathcal{L}}_{EM_{i}}\;.
\end{equation}

Let us first examine the case of pure radiation. Since no driving term
exists for pure radiation, the field $\psi$ does not lose or gain energy
(other than redshifting like $1/a^6$ due to its own pressure) and likewise
the energy for radiation should be defined so that it does not depart from
the usual $1/a^4$ law. Thus for radiation it is clear that a full set of
equations is: 
\begin{eqnarray}
\left( \frac{\dot{a}}{a}\right) ^{2} &=&\frac{1}{3}\left( \widetilde{\rho }%
_{r}+\rho _{\psi }\right) \\
\dot{\widetilde{\rho }}_{r}+4\frac{\dot{a}}{a}\widetilde{\rho }_{r} &=&0 \\
\dot{\rho}_{\psi }+6\frac{\dot{a}}{a}\rho _{\psi } &=&0
\end{eqnarray}
This lends further support to the choice of $\widetilde \rho$ as a suitable
definition for the energy in electromagnetism with varying alpha, as
discussed at the end of last section. The field $\psi$ behaves like a
non-interacting fluid if only radiation is present. Energy should therefore
be defined for radiation (and more generally for electromagnetism) so that
it is also non-interacting in this case.

For other components (including the dark matter) we need equations of state
relating their energy density with their EM Lagrangian content, and this is
not always easy to infer. One possibility is to define parameters: 
\begin{equation}  \label{zeta}
\zeta_i=\frac{\widetilde{\mathcal{L}}^{EM}_i}{\widetilde{\rho}_i}\, .
\end{equation}
For radiation $\zeta_r=0$, but $\zeta_m\neq 0$ for baryonic as well as for
some types of dark matter. The statement that $\zeta_i$ is a constant is
part of the model (and we stress that such a model is \textit{not} the model
employed for matter in~\cite{sbm1}). Dark matter candidates with $\zeta_m<0$
were discussed in~\cite{sbm1}.

A full closed set of cosmological equations for a matter and radiation
universe is therefore: 
\begin{eqnarray}
\left( \frac{\dot{a}}{a}\right) ^{2} &=&\frac{1}{3}\left( \widetilde{\rho }%
_{m}+\widetilde{\rho }_{r}+\rho _{\psi }\right)  \label{fried} \\
\dot{\widetilde{\rho }}_{r}+4\frac{\dot{a}}{a}\widetilde{\rho }_{r} &=&0 \\
\dot{\widetilde{\rho }}_{m}+3\frac{\dot{a}}{a}\widetilde{\rho }_{m} &=&2\dot{%
\psi}\zeta _{m}\widetilde{\rho }_{m} \\
\dot{\rho}_{\psi }+6\frac{\dot{a}}{a}\rho _{\psi } &=&-2\dot{\psi}\zeta _{m}%
\widetilde{\rho }_{m}
\end{eqnarray}%
where $\dot{\psi}$ on the right-hand side of the last two equations can be
written in terms of $\rho _{\psi }$ via (\ref{rhopsi}), to form a closed
system.

\section{Some exact solutions and physical implications}

We now derive some exact solutions with pure matter and scalar kinetic
energy plus matter content (pure radiation is trivial) and discuss the
physical implications.

\subsection{Matter-dominated model}

If $\zeta _{m}$ is indeed a constant we find at once that: 
\begin{equation}
{\tilde{\rho}}_{m}=\frac{Me^{2\zeta _{m}\psi }}{a^{3}}\;,  \label{mat}
\end{equation}%
with $M$ constant. At late times the cosmological equations with zero
radiation density have approximate late-time solutions of the same form as
the earlier BSBM analysis \cite{sbm2}, now with: 
\begin{eqnarray}
a(t) &=&t^{2/3}  \label{asymp} \\
e^{2\zeta _{m}\psi } &=&\frac{\omega_B }{4M\zeta _{m}^{2}\ln (t)}  \notag
\end{eqnarray}%
We observe that in the Friedmann equation this leads to 
\begin{eqnarray*}
{\tilde{\rho}}_{m} &=&\frac{Me^{2\zeta _{m}\psi }}{a^{3}}=\frac{\omega_B }{%
4\zeta _{m}^{2}t^{2}\ln (t)} \\
\rho _{\psi } &=&\omega_B \frac{\dot{\psi}^{2}}{2}=\frac{\omega_B }{8\zeta
_{m}^{2}t^{2}\ln ^{2}(t)}
\end{eqnarray*}%
and so the ${\tilde{\rho}}_{m}$ term dominates increasingly at large $t.$

During any dark-energy dominated era (say driven by a simple cosmological
constant $\Lambda $) the expansion dynamics will approach the de Sitter
evolution $a(t)\propto \exp [t\sqrt{\Lambda /3}]$ and $\psi $ will tend to a
constant value, $\psi _{\infty }$, exponentially rapidly ($\psi \rightarrow
\psi _{\infty }+O(t\exp [-t\sqrt{3\Lambda }])$ as $t\rightarrow \infty $.
Hence in this era the matter density will quickly approach the standard
evolution with ${\tilde{\rho}}_{m}\propto a^{-3}.$

\subsection{Matter with scalar kinetic energy}

If eq. \ref{mat} holds still then the Friedmann equation (\ref{fried})
becomes: 
\begin{equation}
3\left( \frac{\dot{a}}{a}\right) ^{2}=\widetilde{\rho }_{m}+\rho _{\psi }={M}%
\frac{e^{2\zeta _{m}\psi }}{a^{3}}+\frac{1}{2}\omega_B \dot{\psi}^{2}
\label{fr}
\end{equation}%
and the equation for $\psi $ is: 
\begin{equation*}
\ddot{\psi}+3\frac{\dot{a}}{a}\dot{\psi}=-2\zeta _{m}{M}\frac{e^{2\zeta
_{m}\psi }}{a^{3}}\,.
\end{equation*}%
There is a special exact solution of these equations with : 
\begin{eqnarray}
a &=&t^{n}  \label{s1} \\
\psi  &=&A+B\ln (t)  \notag
\end{eqnarray}%
where 
\begin{eqnarray}
3n &=&2(1+B\zeta _{m}),  \label{s2} \\
3n^{2} &=&{M}e^{2\zeta _{m}A}+\frac{1}{2}\omega_B B^{2},  \notag \\
B(1+2B\zeta _{m}) &=&-2\zeta _{m}{M}e^{2\zeta _{m}A},  \notag
\end{eqnarray}%
determine the constants $A,B$ and $n$ in terms of $M,\zeta _{m}$ and $\omega_B 
$. Note that in this solution all the terms in the Friedmann eqn fall as $%
t^{-2}$, specifically: 
\begin{equation*}
{\tilde{\rho}}_{m}\propto \frac{e^{2\zeta _{m}\psi }}{a^{3}}\propto \omega_B 
\dot{\psi}^{2}\propto t^{-2}\;.
\end{equation*}%
However, exact solutions of this simple scaling form are unstable in BSBM 
\cite{bg} and are expected to be so here also, approaching the form \ref%
{asymp} at late times when there is no dark energy term to drive $\psi $ to
a constant value. If $n=1/2$ is chosen in (\ref{s1})-(\ref{s2}) then it is
possible to add a radiation term with $\widetilde{\rho }_{r}\propto
a^{-4}\propto t^{-2}$ to eq. (\ref{fr}) and still obtain an exact solution
of this scaling form.

\subsection{Physical implications}

In contrast with earlier work we find no variations in the radiation
evolution. However, as we have seen in this section, the matter evolution is
changed (including dark matter, depending on the value of $\zeta_m$). This
leads to number of new effects, analogous to those derived based on
variations in the radiation evolution. The main implications are as follows:

\begin{itemize}
\item 
The density of non-relativistic matter which, like weakly interacting
CDM, does not couple to $\psi $ or to electric charge (i.e. $\zeta _{i}=0$),
will scale with expansion as $\rho _{cdm}\propto a^{-3}$. Therefore, in the
general asymptotic matter-dominated solution \ref{asymp} it falls off at a
different rate to that of the electrically coupled matter density, ${\tilde{%
\rho}}_{m}$, and we will have 
\begin{equation*}
\frac{{\tilde{\rho}}_{m}}{\rho _{cdm}}\propto e^{2\zeta _{m}\psi }\propto
1/\ln (t)
\end{equation*}

\item The entropy per baryon is no longer constant during adiabatic expansion.
Although $\widetilde{\rho }_{r}^{3/4}/\rho _{cdm}$ remains constant during
the expansion (and the ratio of the neutrino to photon temperatures will be
constant), the combination $\widetilde{\rho }_{r}^{3/4}/{\tilde{\rho}}%
_{m}\propto e^{-2\zeta _{m}\psi }\propto \ln (t)$ does not remain constant
during the matter-dominated era. Some care must therefore be exercised in
comparing constraints on the entropy per baryon at late times with those
deduced from predictions of the primordial deuterium abundance and
observations.
\end{itemize}

\section{Conclusion}

The absence of dispersion in the free propagation of electromagnetic waves
implies that in BSBM there is no extra shift in the frequency of photons
(other than the usual effect of the cosmological expansion). For a
black-body in free flight this means that the temperature-redshift relation $%
T_{\gamma }(z)=T_{\gamma 0}(1+z)$ is not modified. The spectrum remains
Planckian when the universe expands isotropically and homogeneously, and its
energy density is still given by $\rho _{\gamma }\propto T_{\gamma
}^{4}\propto a^{-4}$. Thus, nothing changes in the temperature-redshift
relation for a black body in free flight in BSBM, as we have shown
explicitly in this paper. This conclusion has been known for a while in the
context of string cosmology and elsewhere~\cite%
{Kaloper:1991mq,Damour:1992kf,Damour:1993id,Wetterich:2002ic,Olive:2001vz},
and in fact is more general than proved here: no energy exchange occurs with
a pure radiation system for any ``multiplicative'' theory (i.e. one of the
dilaton type).

This conclusion remains also true for a coupled, thermalized black body. In
fact this is implied by the Kirchhoff-Clausius law, which states that
\textquotedblleft the rate at which a body emits heat radiation is inversely
proportional to the square of the speed at which the radiation propagates in
the medium in which the body is immersed\textquotedblright\ (e.g. \cite{book}%
). If the medium is non-dispersive, as is the case with BSBM, then the
Planck law and Stefan-Boltzmann law, receive no direct corrections due to a
varying $\alpha $. We still have $\rho \propto 1/a^{4}$ (since $p=1/3\rho $)
and $\rho \propto T_{\gamma }^{4}$, so $T_{\gamma }\propto 1/a$ remains
true. Even though this conclusion seems supported by copious past
literature, it remains controversial in some quarters~\cite{mar1,mar2,hees}
(in part due to a misunderstanding for which~\cite{sbm1} may be partly
responsible). Clearly there are many possible definitions for the energy in
a dielectric, but the one responsible for a conservation law is the most
appropriate. Our sections on cosmological applications should have clarified
the matter.

At first it might appear we have derived a negative result. Since $\rho
\propto 1/a^{4}$ and $\rho \propto T_{\gamma }^{4}$ we may conclude that $%
T_{\gamma }\propto 1/a$ remains true. Obviously the field $\phi $ may still
affect the function $a(t)$, due to its presence in the Friedman equations,
but the temperature-redshift relation remains unmodified and astronomical
observations of $T_{\gamma }(z)$ at $z>0$ will place no new constraints on
the theory. However there are non-trivial effects in this theory because,
even though the radiation evolution is unmodified, the matter (baryonic and
possible dark) evolution is modified. This may lead to changes in the photon
to baryon ratio, as well as the ratios of the various dark matter components.

In a future publication we will develop further the analogy between BSBM and
a dielectric medium so as to generate generalizations of this theory. We
will find that if we are prepared to break Lorentz invariance, then more
interesting phenomenology will arise. The effects of modified dispersion
relations in a different setting have already been studied (see, e.g. \cite%
{steph}). Similar generalizations of conventional dielectrics could be set
up. Even without breaking Lorentz symmetry it is possible to construct
dielectric media which differ from the one representing BSBM theory, with
interesting phenomenology. If we are prepared to break Lorentz invariance, a
wide class of varying speed of light theories will follow \cite{vslrev}.

\section*{Acknowledgments}

We acknowledge STFC for consolidated grant support and thank Aur\'{e}lien
Hees, Julien Larena, Olivier Minazzoli and Alec Graham for helpful
discussions.

\bibliographystyle{plain}
\bibliography{refs-varalpha}

\end{document}